\documentclass[aps,prl,final,twocolumn,letterpaper]{revtex4}

\usepackage{graphicx}   
\usepackage{import}                         
\usepackage{epstopdf}
\usepackage{amsmath} 
\usepackage{float} 
\usepackage{bm}
\usepackage{amssymb}
\usepackage{quotes}
\usepackage{indentfirst}
\usepackage{color}
\usepackage{transparent}
\usepackage{dcolumn}
\usepackage{braket}
\usepackage{multirow}
\usepackage{cancel} 
\usepackage{mdframed}
\usepackage{color}
\usepackage{bm}
\usepackage{dsfont}
\usepackage{slashed}
\usepackage{soul, color} 
\soulregister\ref{7}  
\soulregister\cite{7} 
\renewcommand{\st}[1]{}

\usepackage{xr}

\makeatletter
\newcommand*{\addFileDependency}[1]{
  \typeout{(#1)}
  \@addtofilelist{#1}
  \IfFileExists{#1}{}{\typeout{No file #1.}}
}
\makeatother

\usepackage{textcomp} 
\usepackage{xifthen}
\usepackage{xcolor}
\usepackage{etoolbox}
\newboolean{togglechanges} 

\setboolean{togglechanges}{false}

\newcommand{\comment}[1]{\ifbool{togglechanges}
    {#1}  
    {\textcolor{blue}{#1}}}

\usepackage{bibentry}

\usepackage{graphicx}
\usepackage{dcolumn}
\usepackage{bm}

\usepackage{textcomp} 

\begin{document}
\rmfamily

\title{Measuring, processing, and generating partially coherent light with self-configuring optics}

\author{Charles~Roques-Carmes$^{1}$}
\email{chrc@stanford.edu}
\author{Shanhui~Fan$^{1}$}
\author{David~A.~B.~Miller$^{1}$}

\affiliation{$^{1}$ E. L. Ginzton Laboratory, Stanford University, 348 Via Pueblo, Stanford, CA 94305\looseness=-1}



\clearpage

\renewcommand{\sp}{\sigma_+}
\newcommand{\pbit}{$p$-bit}
\newcommand{\pbits}{$p$-bits}
\newcommand{\sm}{\sigma_-}

\vspace*{-2em}



\begin{abstract}
Optical phenomena always display some degree of partial coherence between their respective degrees of freedom. Partial coherence is of particular interest in multimodal systems, where classical and quantum correlations between spatial, polarization, and spectral degrees of freedom can lead to fascinating phenomena (e.g., entanglement) and be leveraged for advanced imaging and sensing modalities (e.g., in hyperspectral, polarization, and ghost imaging). Here, we present a universal method to analyze, process, and generate spatially partially coherent light in multimode systems by using self-configuring optical networks. Our method relies on cascaded self-configuring layers whose average power outputs are sequentially optimized. Once optimized, the network separates the input light into its mutually incoherent components, which is formally equivalent to a diagonalization of the input density matrix. We illustrate our method with arrays of Mach-Zehnder interferometers and show how this method can be used to perform partially coherent environmental light sensing, generation of multimode partially coherent light with arbitrary coherency matrices, and unscrambling of quantum optical mixtures. We provide guidelines for the experimental realization of this method, paving the way for self-configuring photonic devices that can automatically learn optimal modal representations of partially coherent light fields. 
\end{abstract}

\maketitle

\section*{Introduction} 

In optics and photonics, partially coherent light is the norm rather than the exception and accounts for emission processes in stars, LEDs, thermal emitters, photovoltaics, luminescent and scintillating materials, as well as natural light for sensing the environment~\cite{goodman2015statistical}. The partial coherence of light naturally emerges in various physical phenomena, such as light propagation in turbulent media and astronomy~\cite{korotkova2020applications}. Partially coherent light is also used in advanced imaging, sensing, and communication modalities, such as optical coherence tomography, ghost imaging, stellar interferometry, and low-power optical trapping, to only name a few~\cite{korotkova2020applications}. Partial coherence describes statistical correlations between degrees of freedom of a light field (such as spatial, spectral, polarization, etc.)~\cite{wolf2003unified, de2017unified, korotkova2020applications}. This general description is particularly relevant in understanding phenomena that involve coupled degrees of freedom, such as polarization (meta)optics~\cite{hasman2005space, mueller2017metasurface} and imaging~\cite{rubin2019matrix}, cross-spectral purity~\cite{mandel1961concept}, cylindrical vector beams~\cite{zhan2009cylindrical}, and ``classically entangled'' photonic states~\cite{kagalwala2013bell}.

The coherency matrix $\rho$~\cite{zhang2019scattering, goodman2015statistical} (or its quantum optical analogue, the density matrix~\cite{nielsen2010quantum}) is generally used to characterize such partial coherence over arbitrary channels of a photonic system. Of particular interest is the basis of so-called ``natural modes''~\cite{wolf1982new, wolf1986new, withington1998modal}. We can express any spatially partially coherent optical field near some wavelength as a linear superposition of these modes, which have the important physical property that they are mutually incoherent (i.e., completely uncorrelated). Equivalently, any spatially partially coherent field can be decomposed into orthogonal and mutually incoherent parts. This decomposition is mathematically equivalent to finding the basis that diagonalizes the matrix $\rho$~\cite{wolf1982new, wolf1986new, withington1998modal}. Methods to reconstruct $\rho$ for few polarization-spatial channels have been demonstrated via projective measurements (e.g., for 4$\times$4 polarization$\times$spatial degrees of freedom~\cite{kagalwala2015optical}). Despite the ubiquity of partial coherence in optical phenomena, there is no general, scalable method to measure $\rho$, nor apparently so far any physical method that separates it into its mutually incoherent parts. 

Meshes of Mach-Zehnder interferometers (MZIs)~\cite{bogaerts_programmable_2020} have proven very effective at manipulating~\cite{milanizadeh2021coherent} and measuring~\cite{MillerAnalyze2020} coherent multimode light. MZI meshes have been used to implement inference~\cite{shen2017deep} and training~\cite{pai2023experimentally} in optical neural networks, heuristic algorithms for combinatorial optimization~\cite{prabhu2020accelerating}, simulation of quantum transport~\cite{harris2017quantum}, free space optical control~\cite{milanizadeh2021coherent}, and universal linear optics~\cite{miller2013self, carolan2015universal}. Central to these works is the fact that MZI meshes are universal linear photonic processors~\cite{miller2013self}. Specifically, self-configuring MZI networks can automatically learn unitary operators for coherent light processing~\cite{miller2013self, miller2013self2} and establish optimal communication channels~\cite{miller2013establishing, seyedinnavadeh_determining_2023}. However, the existing literature on MZI meshes predominantly concentrates on coherent light processing, largely overlooking the expansive potential in processing and analyzing incoherent or partially coherent multimode light. 

Here, we propose a general method using self-configuring optics -- ``partially coherent light analyzers'' (PCLA) -- to fully measure the coherency matrix of partially coherent light near some wavelength; this method additionally separates the light into its mutually incoherent orthogonal components, whose powers appear separately in the output waveguides. Our method performs sequential power optimization over the $N$ output channels of a self-configuring network, thereby learning the coherency matrix eigenvectors and eigenvalues. In this process, the unitary network is then also implementing the linear transform that diagonalizes the coherency matrix. If we separately calibrate the network~\cite{MillerAnalyze2020}, we can deduce this diagonalizing transformation (and hence the eigenvectors) from the resulting network settings by simple arithmetic. Together with measurements of the relative output powers, which give the matrix eigenvalues, this process therefore measures this matrix. 

We illustrate our method in three distinct settings: (1) analyzing partially coherent environmental light from a scene; (2) generating partially coherent light with an arbitrary coherency matrix; (3) analyzing incoherent mixtures of single photons on an integrated photonic network. Our method therefore paves the way to full characterization and processing of partially coherent light, addressing significant untapped opportunities in fields such as environmental and astronomical sensing, quantum optics, and advanced imaging, in each of which partial coherence plays a fundamental role.

\begin{figure*}
\centering
\vspace{-0.2cm}
  \includegraphics[scale=0.75]{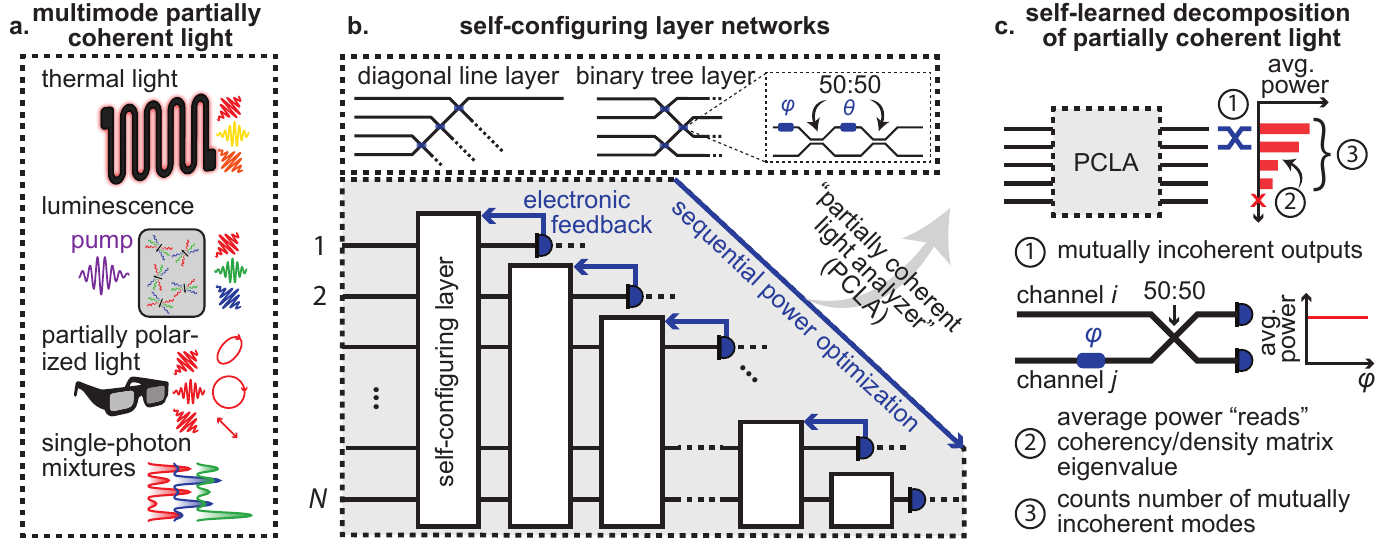}
    \caption{\small \textbf{Measuring and processing partially coherent light with self-configuring optics.} \textbf{a.} Partial coherence of light is observed in many photonic systems: between spatial modes in thermal light emission or luminescent materials (pumped by optical light or high-energy particles); partially polarized light in environmental sensing; or as incoherent mixtures of pure states in quantum optics. \textbf{b.} Partial coherent light analyzer (PCLA): self-configuring networks can automatically analyze these different forms of multimode partially coherent light near some wavelength of interest. Input multimode partially coherent light is coupled into $N$ waveguides that then feed a cascade of self-configuring layers (e.g., each a diagonal line or binary tree of Mach-Zehnder interferometers (MZIs)). These layers then learn a decomposition of the corresponding density matrix $\rho$ via a sequential, layer-by-layer power optimization method relying on measurement and feedback. Each node of the array is a $2 \times 2$ MZI.  \textbf{c.} The learned decomposition separates the mutually incoherent modes (eigenvectors of $\rho$) to generate outputs (1) from the ``top'' waveguide of each self-configuring layer. The resulting network settings give the eigenvectors. The output power of each such waveguide (2) gives the corresponding eigenvalue of $\rho$; and (3) the number of output ports with non-zero power corresponds to the number of mutually incoherent modes (up to a maximum of \textit{N}). }
    \label{fig:concept}
    \vspace{-0.3cm}
\end{figure*}

\section*{Self-learning partially coherent light analyzers (PCLA)} 

We first describe the physics and learning procedure of PCLA in processing partially coherent light. Our approach can in principle process partial coherence over many spatial degrees of freedom of a light field and in various settings, with some examples shown in Fig.~\ref{fig:concept}a. We collect the input light into $N$ spatial ``channels'' or waveguides into the PCLA, using grating or other input couplers. Polarization splitting couplers that route different input polarizations to waves in the same polarization in different waveguides would add the ability simultaneously to process polarization degrees of freedom also~\cite{miller2013self}. 

Our PCLAs consist of a cascade of up to \textit{N} self-configuring layers, such as diagonal lines~\cite{miller2013self2, miller2013self} (resulting in a triangular mesh~\cite{reck1994experimental}), binary tree layers~\cite{miller2013self2, MillerAnalyze2020}, or hybrid architectures~\cite{MillerAnalyze2020, pai2022scalable}, all constructed from 2$\times$2 programmable interferometer blocks. Self-configuring layers can be defined topologically as ones in which there is one and only one path through these blocks from the ``top'' output (Fig.~\ref{fig:concept}b) of the layer to each input to the layer~\cite{MillerAnalyze2020}. 

For concreteness, we consider integrated self-configuring layers, with the 2$\times$2 blocks implemented using integrated MZIs. Such MZIs are made from two phase shifters ($\theta, \phi$) and two 50:50 directional couplers (Fig.~\ref{fig:concept}b)~\cite{miller2013self2,MillerAnalyze2020,miller2015perfect,wilkes201660}. Each layer has a single (``top'') output whose power is measured with a photodetector. That measurement is used to update the settings of that layer via electronic feedback. The photodetector could be an external or integrated photodiode, and could also be designed just to sample a sufficient amount of power during measurement, leaving the majority of the separated output power for other purposes. Specifically, each layer optimizes (e.g., maximizes) the power at each detector by tuning the parameters (e.g., phase shifters) of that single layer. In the self-configuring geometry, the power output of a given layer is independent of the parameters of all subsequent layers, thereby reducing the number of degrees of freedom for each subsequent power optimization. The power optimization is sequential: the power output of the first layer is first maximized, then that of the second, and so forth.

Once the sequential power optimization has converged, the PCLA has learned a modal representation of the spatially partially coherent input light field corresponding to mutually incoherent modes~\cite{wolf1982new, wolf1986new, withington1998modal} (see Fig.~\ref{fig:concept}c). These mutually incoherent modes do not produce interference patterns when mixed with a tunable phase (a feature that can be further checked experimentally with an analyzer network after the PCLA, as discussed later in this paper). In the process, the PCLA has learned the coherency matrix eigenvectors, which can then be deduced directly from the resulting settings of the (calibrated~\cite{MillerAnalyze2020}) network elements. The corresponding eigenvalues $\lambda_i$ can be measured by reading out the average values of the output powers. Additionally, the number of output ports with non-zero average power corresponds to the number of such mutually incoherent modes, or the rank of the coherency matrix.

We now describe the PCLA learning procedure. Let us denote $\rho_\text{in}$ as the coherency matrix of the input field. The coherency matrix is Hermitian semi-positive and can therefore be diagonalized as:
\begin{equation}
    \rho_\text{in} = U D U^\dagger,
    \label{eq:density-transformation}
\end{equation}
where $U$ is the orthogonal basis of mutually incoherent eigenmodes and $D$ a positive diagonal matrix corresponding to the average power in each mode $\lambda_i \geq 0$. Characterizing $\rho_\text{in}$ entails measuring the unitary operator $U$ and the eigenvalues $\lambda_i$. 

A linear operation $U_\text{PCLA}$ on these channels transforms the coherency matrix as : $\rho_\text{out} = U_\text{PCLA} \rho_\text{in} U_\text{PCLA}^\dagger$~\cite{zhang2019scattering} (where $\rho_\text{out}$ is the coherency matrix of the network output $y$). Each step of the algorithm consists in the maximization of the ensemble averaged power at the output port of one of the self-configuring layers. At step $k$, the network optimization is the following:
\begin{equation}
    \underset{S_k}{\text{max}} \left( \rho_{\text{out}}\right)_{kk} = \lambda_{k},
\end{equation}
where $\lambda_k$ is the $k$-th largest eigenvalue of $\rho_\text{in}$ (ordered such that $\lambda_1\geq\ldots\geq\lambda_{N}$), and $S_k$ is set of tunable parameters (phases) in the $k$-th self-configuring layer, corresponding to a set of MZI denoted $M_k$. This equality is a direct consequence of the min-max or variational theorem of linear algebra, whose conditions are naturally enforced in self-configuring networks due to the mutual orthogonality of the self-configuring layers~\cite{miller2013self}. We can then optimize the network settings sequentially for one layer at a time, and the relative power at output node $k$ gives $\lambda_k$.

The PCLA therefore ``diagonalizes'' the coherency matrix $\rho_\text{in}$, such that $U_\text{PCLA} = U^\dagger$. Consequently, reading out the network parameters and output powers fully characterizes the coherency matrix. More details of the proof can be found in Section~S1 of the Supplementary Materials (SM). In the following, we illustrate this method in several settings where partial coherence of light is essential. 

Once configured, one can know the values of the phase delays in the phase shifters by reading the applied voltages (or other control variables). Approaches to the necessary calibration of the phase shifters include progressive methods presuming 50:50 beamsplitters~\cite{MillerAnalyze2020, miller2017setting}, methods of setting up and calibrating ``perfect'' meshes even when the fabricated beamsplitters are not 50:50~\cite{miller2015perfect}, and approximate methods based on global optimization~\cite{butow2022spatially}. Therefore, once the PCLA has performed the sequential power optimization, we can read off these voltages or control values and deduce exactly the unitary matrix $U_\text{PCLA}$ represented by the mesh~\cite{MillerAnalyze2020}.

\section*{Environmental light processing with self-configuring Mach-Zehnder interferometer arrays}

\begin{figure}
\centering
\vspace{-0.2cm}
  \includegraphics[scale=0.9]{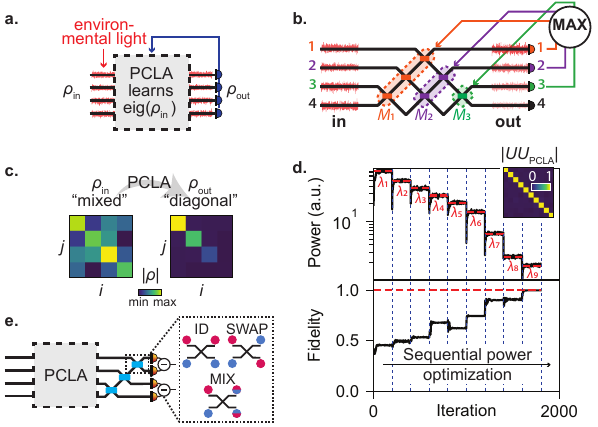}
    \caption{\small \textbf{Environmental light processing with self-configuring Mach-Zehnder interferometer arrays.} \textbf{a.} A PCLA can process environmental light input and learn its decomposition into mutually incoherent modes (eigenvalue decomposition of input density matrix $\rho_\text{in}$).  \textbf{b.} An incident wavefront of partially coherent waves is sent through a triangular array of MZI. A sequential power optimization learns a decomposition of the incident wavefront into mutually incoherent modes. \textbf{c.} 4-mode example input and output coherency matrices, the latter resulting from the sequential power optimization. \textbf{d.} Top: The power optimization routine sequentially maximizes the power at each output channel (with channel index $i$ from $N=1$ to $4$), mapping each output sequentially to the corresponding coherency matrix eigenvalue ($\lambda_i$). Inset: Matrix product showing that $U_\text{PCLA}$ learns $U^\dagger$ (up to a diagonal matrix of phases). Bottom: Resulting fidelity over power optimization iteration. \textbf{e.} Analyzer circuit, consisting in a single layer of nodes that can be configured in either one of three gates: identity, swap, and Hadamard gates (mixing). The output signals are then analyzed via balanced homodyne measurements.}
    \label{fig:environmental}
    \vspace{-0.3cm}
\end{figure}

We now show how PCLAs can be used to analyze and process partially coherent light fields impinging on the PCLA from a scene, as shown in Fig.~\ref{fig:environmental}a. As a practical matter, the behavior of circuit components such as input couplers, waveguide beamsplitters and phase shifters will depend on the wavelength to some degree, but we presume that the spectral bandwidth of the input light is narrow enough or has been sufficiently filtered that we can approximately neglect such dependence for our discussion. The interferometer meshes themselves can be constructed with path lengths that are all essentially equal for all interfering components~\cite{MillerAnalyze2020}, so the behavior of the meshes is otherwise essentially independent of wavelength. 

We consider $N$ ``channels'' of input light, whose fluctuating amplitudes are denoted by an $N$-dimensional vector $x$. The partial coherence of these channels is described by the coherency matrix $\rho_\text{in}$~\cite{goodman2015statistical}, such that $\left( \rho_\text{in}\right)_{ij} = \langle x_i x_j^*\rangle$, where $\langle\cdot \rangle$ denotes ensemble averaging (e.g., time averaging if we presume stationary ergodic fields~\cite{goodman2015statistical}). For illustrative purposes, each node of the network in a 4-channel triangular array example (Fig.~\ref{fig:environmental}b) is labelled with the corresponding output port optimization color (with $M_1$ shown in orange, $M_2$ in purple, and $M_3$ in green, respectively). 

We demonstrate the validity of our approach with numerical experiments in Fig.~\ref{fig:environmental}c-d with a 10-channel fluctuating input field, simulating light propagation with fluctuating amplitudes through a triangular array. As the power optimization is carried out, each channel's output power gives the corresponding eigenvalue of $\rho_\text{in}$. The corresponding unitary fidelity (defined as $F=\langle |U_\text{PCLA} U|,~\text{Id}\rangle_\text{HS}$~\cite{prabhu2020accelerating}, where $\langle\cdot\rangle_\text{HS}$ is the Hilbert-Schmidt dot product and Id is the identity matrix) increases throughout the power optimization and reaches values $>0.99$ after convergence, thereby showing the PCLA learns the spectral representation of the coherency matrix with great accuracy. 

Once configured, the fields in different output channels of this network should be mutually incoherent; if we then attempt to interfere each pair of outputs, we should see no interference between them as the relative phase of those outputs is varied. To test such mutual incoherence, one can use an additional output analyzer layer of MZIs, as in Fig.~\ref{fig:environmental}e, after the coherency diagonalization circuit $U_\text{PCLA} = U^\dagger$. To interfere any two outputs, the MZI nodes can be appropriately configured as (1) identity; (2) swap; (3) or mix (i.e., 50:50 splitter, as in Hadamard gates), shown in Fig.~\ref{fig:environmental}e, onto an output photodetector. Scanning the relative input phase using the analyzer input phase shifters should then produce no interference fringes (see Fig.~\ref{fig:concept}c), which is equivalent to performing balanced homodyne measurements, yielding a zero-mean power. Details of the parameters and methods used in this numerical experiment can be found in Section~S2 of the SM.

\begin{figure}
\centering
\vspace{-0.2cm}
  \includegraphics[scale=0.9]{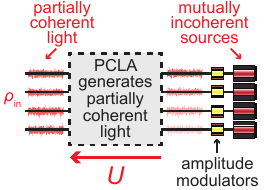}
    \caption{\small \textbf{Generation of multimode partially coherent light with PCLA.} To generate multimode partially coherent light described by coherency matrix $\rho_\text{in}$, the PCLA is run backwards and illuminated with mutually incoherent sources. Their amplitude is set, for example with modulators as shown, so that their average power matches the eigenvalues $\lambda_i$ of the coherency matrix.}
    \label{fig:generation}
    \vspace{-0.3cm}
\end{figure}

Incidentally, PCLAs can also be used to \textit{generate} multimode partially coherent light described by an arbitrary coherency matrix. Running the self-configuring network of this section in the backwards direction, as shown in Fig.~\ref{fig:generation}, we illuminate its output ports with mutually incoherent sources with average powers corresponding to the desired eigenvalues of the coherency matrix $\lambda_i$. The resulting coherency matrix emerging ``backwards'' on the input side is that of partially coherent light and described as in Eq.~(\ref{eq:density-transformation}), choosing $U_\text{PCLA} = U^\dagger$. Knowing the natural mode decomposition of the coherency matrix of a desired partially coherent light field of interest, one can therefore use PCLAs to generate such a light field using mutually incoherent sources of variable power. 

\section*{Processing incoherent mixtures of delocalized single photons with PCLA} 

\begin{figure}
\centering
  \includegraphics[scale=0.96]{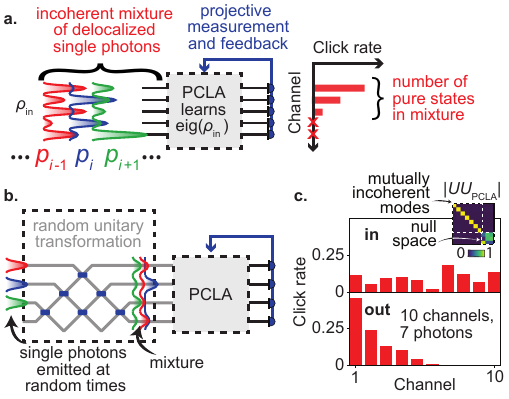}
    \caption{\small \textbf{Processing incoherent mixtures of delocalized single photons with PCLA.} \textbf{a.} An incoherent mixture of delocalized single photons (described by coherency matrix $\rho_\text{in}$) is sent through a PCLA. The result of projective measurements is used to sequentially optimize the power at each output mode. The result is a modal decomposition into mutually incoherent modes, where the number of pure quantum states in the mixture corresponds to the number of output modes with non-zero average power. \textbf{b.} Experimental proposal, where single photons emitted at random times (therefore mutually incoherent) are mixed through a random unitary transformation and subsequently analyzed by the PCLA. \textbf{c.} 10-mode example (with a total of 7 single photons): the output clicks resulting from the sequential power optimization are ordered and map to the eigenvalues of $\rho_\text{in}$. Inset: Matrix product showing that $U_\text{PCLA}$ learns eigenvectors of $\rho_\text{in}$ (up to a diagonal matrix of phases).}
    \label{fig:quantum}
    \vspace{-0.3cm}
\end{figure}

We now further generalize our method to analyzing incoherence in quantum optical systems, generally described by a density matrix $\rho_\text{in}$, and illustrated in an integrated photonic network where single photons propagate~(Fig.~\ref{fig:quantum}a). 

We consider incoherent mixtures of single photons delocalized over $N$ waveguide ports (Fig.~\ref{fig:quantum}a). The input mixed state is described by $\rho_\text{in} = \sum_i p_i \ket{\psi_i} \bra{\psi_i}$, with $0<p_i <1$. Note, as is typical with mixed states, that the different $\ket{\psi_i}$ and the corresponding optical waves arriving at the PCLA need not be orthogonal to one another. The PCLA imparts a unitary transformation to the wavefunction $\ket{\psi_\text{out}}=U_\text{PCLA}\ket{\psi_\text{in}}$, which corresponds to the following operation on the density matrix, similar to that on the coherency matrix in the previous examples~\cite{nielsen2010quantum}: $\rho_\text{out} = U_\text{PCLA} \rho_\text{in} U_\text{PCLA}^\dagger$. Detectors on the PCLA output measure ``clicks'' corresponding to single photons, and the average power at a given output channel $k$ is given by $\left(\rho_\text{out}\right)_{kk}$. Therefore, a sequential power optimization analogous to that of the previous sections can be carried out to analyze incoherence of this quantum optical system.

The nature of this incoherent mixture of single photons is responsible for stochastic fluctuations of the output power. Specifically, stochasticity arises from two sources: (1) the classical incoherent mixture, from which each pure state $\ket{\psi_i}$ can be ``picked'' with probability $p_i$; (2) projective quantum measurements on the output, with probability of clicking at output port $k$ given by $|\braket{k|U_\text{PCLA}|\psi_i}|^2$. Both effects can be modeled by a categorical distribution, and more details on our numerical implementation can be found in the SM, Section~S3. 

Both sources of randomness are simulated in the thought experiment shown in Fig.~\ref{fig:quantum}b, where a random unitary transformation is first imparted to single photons emitted at random times (thereby providing incoherence of the mixture on the output). In this example, we propagate a 7-photon mixture through the PCLA and perform a sequential power optimization as described in the previous sections. While the input density matrix was mixed, the output of the PCLA after optimization is ordered with decreasing mean power. The number of channels with non-zero mean power corresponds to the number of pure states in the mixture, and the PCLA has learned the modal (diagonal) representation of the density matrix that outputs mutually incoherent modes.

\section*{Discussion} 
We further discuss potential applications of our method and experimental considerations for their realization. 

We have shown that PCLAs, which consist of self-configuring networks with sequentially optimized power outputs, can be utilized to automatically analyze the classical and quantum partial coherence of multimode optical light fields.
Quite generally, our methods highlight the interplay between coherence and multimodal coupling in the analysis of partially coherent light fields. 


Our method also displays a few distinctive advantages compared to tomographic reconstruction of the coherence function~\cite{kagalwala2015optical}. Once the PCLA's learning algorithm has converged, it will naturally separate the input light field into its mutually incoherent components. To put it differently, the PCLA acts as a \textit{lossless} ``unscrambler'' of partially coherent light into its mutually incoherent parts. Further connections of our method to other modal representations of partially coherent light fields are discussed in Section~S4 of the SM.

In our numerical experiments, gradients of the time-averaged output powers were calculated using automatic differentiation and optimized with stochastic gradient descent~\cite{kingma2014adam}. In experimental implementations, various gradient calculation or measurement techniques could be used, such as \textit{in situ} back-propagation~\cite{hughes2018training} or dithering~\cite{milanizadeh2021coherent,seyedinnavadeh_determining_2023}. Alternatively, methods such as physical gradient descent~\cite{wright2022deep} or gradient-free physical gradients~\cite{momeni2023backpropagation} could be used. 

In conclusion, we have shown that self-configuring photonic networks, such as triangular arrays of MZIs, can automatically learn and measure the coherency matrix of a multimodal light field across $N$ channels. Our method generalizes to quantum optical systems, as long as enough degrees of freedom are available to implement arbitrary unitary transformations on their Hilbert space. We envision that this method will be experimentally relevant in processing, imaging, and analyzing the classical and quantum coherence of light and matter in all systems and applications where spatially partially coherent light emission is of importance.


\section{Competing interests}
The authors declare no potential competing financial interests.

\section{Data and code availability statement}
The data and codes that support the plots within this paper and other findings of this study are available from the corresponding authors upon reasonable request. Correspondence and requests for materials should be addressed to C.~R.-C. (chrc@stanford.edu).

\section{Acknowledgements}
The authors would like to thank Philipp Del Hougne, Cheng Guo, Carson Valdez, Annie Kroo, Anna Miller, and Olav Solgaard for stimulating conversations. C.~R.-C. is supported by a Stanford Science Fellowship. S.~F. and D.A.B.~M. acknowledges support by the Air Force Office of Scientific Research (AFOSR, grant FA9550-21-1-0312)

\bibliography{bibliography}

\end{document}